\documentclass[%
reprint,
superscriptaddress,
amsmath,amssymb,
prd,
floatfix,
]{revtex4-2}

\usepackage{graphicx}
\usepackage{dcolumn}
\usepackage{bm}
\usepackage[colorlinks=true,allcolors=blue]{hyperref}
\usepackage{float}

\newcommand{\xeonetwofour}{\textsuperscript{124}Xe}
\newcommand{\xeonetwonine}{\textsuperscript{129}Xe}
\newcommand{\xeonethreeone}{\textsuperscript{131}Xe}

\newcommand{\rnzero}{\textsuperscript{220}Rn}
\newcommand{\rntwo}{\textsuperscript{222}Rn}
\newcommand{\krfive}{\textsuperscript{85}Kr}
\newcommand{\krthree}{\textsuperscript{83m}Kr}
\newcommand{\iodine}{\textsuperscript{125}I}
\newcommand{\lead}{\textsuperscript{214}Pb}
\newcommand{\dec}{2$\nu$ECEC}

\newcommand{\1}[1]{\;\mathrm{#1}}
\newcommand{\gev}{\ensuremath{\mathrm{GeV\!/c^2}}}
\newcommand{\cstwob}{\ensuremath{\mathrm{cS2_b}}}

\begin{document}

\preprint{APS/123-QED}

\title{Search for inelastic scattering of WIMP dark matter in XENON1T}

\newcommand{\bologna}{\affiliation{Department of Physics and Astronomy, University of Bologna and INFN-Bologna, 40126 Bologna, Italy}}
\newcommand{\chicago}{\affiliation{Department of Physics \& Kavli Institute for Cosmological Physics, University of Chicago, Chicago, IL 60637, USA}}
\newcommand{\coimbra}{\affiliation{LIBPhys, Department of Physics, University of Coimbra, 3004-516 Coimbra, Portugal}}
\newcommand{\columbia}{\affiliation{Physics Department, Columbia University, New York, NY 10027, USA}}
\newcommand{\lngs}{\affiliation{INFN-Laboratori Nazionali del Gran Sasso and Gran Sasso Science Institute, 67100 L'Aquila, Italy}}
\newcommand{\mainz}{\affiliation{Institut f\"ur Physik \& Exzellenzcluster PRISMA, Johannes Gutenberg-Universit\"at Mainz, 55099 Mainz, Germany}}
\newcommand{\heidelberg}{\affiliation{Max-Planck-Institut f\"ur Kernphysik, 69117 Heidelberg, Germany}}
\newcommand{\munster}{\affiliation{Institut f\"ur Kernphysik, Westf\"alische Wilhelms-Universit\"at M\"unster, 48149 M\"unster, Germany}}
\newcommand{\nikhef}{\affiliation{Nikhef and the University of Amsterdam, Science Park, 1098XG Amsterdam, Netherlands}}
\newcommand{\nyuad}{\affiliation{New York University Abu Dhabi, Abu Dhabi, United Arab Emirates}}
\newcommand{\purdue}{\affiliation{Department of Physics and Astronomy, Purdue University, West Lafayette, IN 47907, USA}}
\newcommand{\rpi}{\affiliation{Department of Physics, Applied Physics and Astronomy, Rensselaer Polytechnic Institute, Troy, NY 12180, USA}}
\newcommand{\rice}{\affiliation{Department of Physics and Astronomy, Rice University, Houston, TX 77005, USA}}
\newcommand{\stockholm}{\affiliation{Oskar Klein Centre, Department of Physics, Stockholm University, AlbaNova, Stockholm SE-10691, Sweden}}
\newcommand{\subatech}{\affiliation{SUBATECH, IMT Atlantique, CNRS/IN2P3, Universit\'e de Nantes, Nantes 44307, France}}
\newcommand{\torino}{\affiliation{INAF-Astrophysical Observatory of Torino, Department of Physics, University  of  Torino and  INFN-Torino,  10125  Torino,  Italy}}
\newcommand{\ucsd}{\affiliation{Department of Physics, University of California San Diego, La Jolla, CA 92093, USA}}
\newcommand{\wis}{\affiliation{Department of Particle Physics and Astrophysics, Weizmann Institute of Science, Rehovot 7610001, Israel}}
\newcommand{\zurich}{\affiliation{Physik-Institut, University of Z\"urich, 8057  Z\"urich, Switzerland}}
\newcommand{\paris}{\affiliation{LPNHE, Sorbonne Universit\'{e}, Universit\'{e} de Paris, CNRS/IN2P3, Paris, France}}
\newcommand{\freiburg}{\affiliation{Physikalisches Institut, Universit\"at Freiburg, 79104 Freiburg, Germany}}
\newcommand{\lal}{\affiliation{Universit\'{e} Paris-Saclay, CNRS/IN2P3, IJCLab, 91405 Orsay, France}}
\newcommand{\napels}{\affiliation{Department of Physics ``Ettore Pancini'', University of Napoli and INFN-Napoli, 80126 Napoli, Italy}}
\newcommand{\nagoya}{\affiliation{Kobayashi-Maskawa Institute for the Origin of Particles and the Universe, and Institute for Space-Earth Environmental Research, Nagoya University, Furo-cho, Chikusa-ku, Nagoya, Aichi 464-8602, Japan}}
\newcommand{\laquila}{\affiliation{Department of Physics and Chemistry, University of L'Aquila, 67100 L'Aquila, Italy}}
\newcommand{\tokyo}{\affiliation{Kamioka Observatory, Institute for Cosmic Ray Research, and Kavli Institute for the Physics and Mathematics of the Universe (WPI), the University of Tokyo, Higashi-Mozumi, Kamioka, Hida, Gifu 506-1205, Japan}}
\newcommand{\kobe}{\affiliation{Department of Physics, Kobe University, Kobe, Hyogo 657-8501, Japan}}
\newcommand{\ucla}{\affiliation{Physics \& Astronomy Department, University of California, Los Angeles, CA 90095, USA}}
\newcommand{\kit}{\affiliation{Institute for Astroparticle Physics, Karlsruhe Institute of Technology, 76021 Karlsruhe, Germany}}
\newcommand{\tsinghua}{\affiliation{Department of Physics \& Center for High Energy Physics, Tsinghua University, Beijing 100084, China}}
\newcommand{\alsoatferrara}{\affiliation{INFN, Sez. di Ferrara and Dip. di Fisica e Scienze della Terra, Universit\`a di Ferrara, via G. Saragat 1, Edificio C, I-44122 Ferrara (FE), Italy}}
\newcommand{\alsoatsuny}{\affiliation{Simons Center for Geometry and Physics and C. N. Yang Institute for Theoretical Physics, SUNY, Stony Brook, NY, USA}}
\newcommand{\alsoatutrecht}{\affiliation{Institute for Subatomic Physics, Utrecht University, Utrecht, Netherlands}}
\newcommand{\alsoatcoimbrapoli}{\affiliation{Coimbra Polytechnic - ISEC, Coimbra, Portugal}}
\newcommand{\alsoatiarnagoya}{\affiliation{Institute for Advanced Research, Nagoya University, Nagoya, Aichi 464-8601, Japan}}
\author{E.~Aprile}\columbia
\author{J.~Aalbers}\stockholm
\author{F.~Agostini}\bologna
\author{M.~Alfonsi}\mainz
\author{L.~Althueser}\munster
\author{F.~D.~Amaro}\coimbra
\author{S.~Andaloro}\rice
\author{E.~Angelino}\torino
\author{J.~R.~Angevaare}\nikhef
\author{V.~C.~Antochi}\stockholm
% ...

\collaboration{XENON Collaboration}
\email[]{xenon@lngs.infn.it}
\noaffiliation

\date{26th February 2021}

\begin{abstract}
We report the results of a search for the inelastic scattering of weakly interacting massive particles (WIMPs) in the XENON1T dark matter experiment.
Scattering off \xeonetwonine{} is the most sensitive probe of inelastic WIMP interactions, with a signature of a $39.6\1{keV}$ de-excitation photon detected simultaneously with the nuclear recoil.
Using an exposure of 0.83 tonne-years, we find no evidence of inelastic WIMP scattering with a significance of more than 2$\sigma$.
A profile-likelihood ratio analysis is used to set upper limits on the cross-section of WIMP-nucleus interactions.
We exclude new parameter space for WIMPs heavier than $100\1{\gev}$, with the strongest upper limit of $3.3 \times 10^{-39}$ cm${}^2$ for 130 \gev{} WIMPs at 90\% confidence level.
\end{abstract}



\maketitle


\section{Introduction}
\label{sec:intro}
A wealth of astrophysical and cosmological evidence points towards the existence of dark matter \cite{Bertone:2004pz,Aghanim:2018eyx}.
Of the many postulated candidates for dark matter, the weakly interacting massive particle (WIMP) is particularly well-motivated, and would be expected to have directly detectable interactions with baryonic matter~\cite{Roszkowski:2017nbc,Feng:2010gw}.
A wide variety of experiments have searched for such an interaction, but a convincing signal is yet to be observed~\cite{Schumann:2019eaa,Conrad:2017pms,Kahlhoefer:2017dnp}.
In this work, we present a search for the inelastic scattering of WIMPs off nuclei in the XENON1T experiment~\cite{Aprile:2017aty}.

The main focus of direct detection searches is usually the elastic scattering of WIMP dark matter off target nuclei, aiming to detect the $\mathcal{O}(10\1{keV})$ recoiling nucleus~\mbox{\cite{Aprile:2018dbl, Agnese:2017jvy, Akerib:2016vxi, Cui:2017nnn}}.
This work concentrates instead on inelastic scattering, where the target nucleus is left in an excited state after the interaction.
The subsequent de-excitation of the nucleus produces a characteristic $\gamma$-ray which is detected as an additional energy deposit in the detector, alongside the kinetic energy transferred to the nucleus.
The inelastic scattering of dark matter discussed here is not to be confused with so-called inelastic dark matter models, in which the dark matter particle itself can be excited during interactions~\cite{TuckerSmith:2001hy}.

We concentrate on scattering off a particular isotope: \xeonetwonine{}.
It has the lowest-lying excited state of any xenon isotope as well as a relatively high natural abundance of $26\%$, meaning that it dominates the expected rate of inelastic WIMP scattering in xenon-based detectors.
The first excited $3/2^+$ state lies $39.6\1{keV}$ above the $1/2^+$ ground state and has a half-life of $0.97\1{ns}$~\cite{Timar:2014rty}.
The structure functions for inelastic scattering off \xeonetwonine{} were calculated in~\cite{Baudis:2013bba}, and the corresponding recoil spectra are used for this work.
An additional channel exists in \xeonethreeone, with a slightly lower isotopic abundance of $21\%$.
However, because the first excited $1/2^+$ state lies $80.2\1{keV}$ above the $3/2^+$ ground state (half-life of $0.48\1{ns}$)~\cite{Khazov:2006igf}, this channel is suppressed and not considered in this analysis~\cite{McCabe:2015eia}.

Direct detection experiments such as XENON1T tend to focus on searches for elastic WIMP scattering, where they have the highest sensitivity to most interaction models.
This is due to their ability to discriminate nuclear recoil (NR) WIMP signals from electronic recoils (ERs), which constitute the majority of the backgrounds.
As ERs constitute the majority of events in the energy range relevant for this search, the $\gamma$-ray component of the signal for inelastic scattering worsens this background discrimination.
That said, inelastic scattering is not only a complementary search strategy.
An observation of WIMP inelastic scattering would also constrain the properties of the WIMP beyond what is possible with elastic scattering only.
In addition, the inelastic signal would only be detectable for spin-dependent interactions, whereas an elastic scattering signal could be detected for both spin-dependent and spin-independent interactions~\cite{McCabe:2015eia}.
Finally, there are some interaction models for which the inelastic channel is more sensitive~\cite{Arcadi:2019hrw}.

\section{The XENON1T detector}
\label{sec:xenon1t}

Located in the Gran Sasso National Laboratory (LNGS) in central Italy, the XENON1T detector was a cylindrical dual-phase (liquid/gas) xenon time projection chamber (TPC)~\cite{Aprile:2017aty}, operated between 2016 and 2018.
The active target was $96\1{cm}$ in diameter and $97\1{cm}$ in length, enclosing $2\1{t}$ of ultra-pure xenon.
This was viewed from above and below by two arrays of 127 and 121 Hamamatsu R11410-21 photomultipler tubes (PMTs), respectively~\cite{Barrow:2016doe,Aprile:2015lha}.
An additional $1.2\1{t}$ of xenon lay between the TPC and the inner cryostat vessel, providing a layer of passive shielding.
Three electrodes, the cathode located near the bottom of the TPC, the gate just below the liquid surface, and the anode just above it, established the electric fields necessary for the detector operation.

The TPC itself was housed in a vacuum-insulated cryostat, which in turn was housed in a large tank $\sim10\1{m}$ in diameter and height.
This tank was filled with deionised water and instrumented with 86~Hamamatsu R5912 PMTs as an active water-Cherenkov muon veto~\cite{Aprile:2014zvw}.
A service building adjacent to the water tank holds supporting infrastructure elements such as the cryogenic system, xenon purification, distillation~\cite{Aprile:2016xhi}, and storage, data acquisition~\cite{Aprile:2019cee}, and slow control.
The materials forming the cryostats and all detector components were selected after a rigorous material screening campaign to ensure high radiopurity~\cite{Aprile:2017ilq,Aprile:2020vmn}.

When a particle interacts in the liquid target, scintillation light is produced and xenon atoms are ionized.
The scintillation light is detected by the PMTs and forms the S1 signal.
The drift field between the cathode and gate drifts the liberated electrons up towards the liquid surface, where the much stronger extraction field between the gate and anode extracts the electrons into the gas and causes energetic collisions with the xenon atoms which results in further scintillation.
This second, amplified signal constitutes the ionisation signal or S2, which is proportional to the number of extracted electrons.
The full reconstruction of the interaction vertex is achieved through the combination of the time between S1 and S2 signals (providing the $z$ coordinate) and the illumination pattern on the top PMT array (providing the $(x,y)$ coordinates).
The energy deposited in the interaction is reconstructed as a linear combination of the scintillation and ionisation signals~\cite{Aprile:2020yad}.

\section{Data analysis}
\label{sec:analysis}

An inelastic scattering event's signature comes from two distinct energy depositions: the NR itself and a subsequent ER from the $39.6\1{keV}$ $\gamma$-ray produced by nuclear de-excitation.
We assume that the two energy depositions are effectively simultaneous, since the $0.97\1{ns}$ half-life of the excited state is much shorter than the $10\1{ns}$ time resolution of the data acquisition.
The mean photon-absorption length at this energy in liquid xenon is $150\1{\mu m}$~\cite{NIST:XCOM}.
This is much longer than the electron-ion recombination length scale of $4.6\1{\mu m}$~\cite{Mozumder_1995}.
We therefore assume that the liquid xenon response to each deposition is unaffected by the other and that they can therefore be treated as independent.
We also note that a $150\1{\mu m}$ difference in the depth of the two interactions corresponds to a difference in arrival time for the two S2s on the order of $0.1\1{\mu s}$.
As this is substantially smaller than the width of a typical S2 signal ($\mathcal{O}(1\1{\mu s})$), we expect inelastic scattering events' S2s to appear like those from a single energy deposition.

\subsection{Event selection}
\label{sec:eventsel}

Data selection follows~\cite{Aprile:2018dbl} and~\cite{XENON:2019dti} very closely, with extended details and discussion in~\cite{Aprile:2019bbb}.
Data were selected from the period between 2 February 2017 and 8 February 2018, when the detector conditions were stable and the drift field was constant.
Because calibration signals from \krthree{} are very close to the signal region of interest in this search, data taken shortly after calibration periods using this isotope were removed, resulting in a total of $234.2\1{days}$ of livetime.

From this initial data selection, additional quality requirements are imposed.
To limit possible bias, all events near the signal region were blinded for this work prior to the start of this analysis.
The same fiducial volume was used as in~\cite{Aprile:2018dbl}, containing $1316\1{kg}$ of xenon.
Inside this volume, events are selected that are consistent with a well-reconstructed single-site interaction and do not occur shortly after a high-energy event, where the delayed extraction of electrons~\cite{Akerib:2020jud} produces additional S2 signals that can adversely affect the event reconstruction.
Events are required to have an energy deposition between $15$ and $71\1{keV}$, which forms the region of interest for this analysis, based on the expected signal and backgrounds (see sections~\ref{sec:signal} and~\ref{sec:backgrounds}).
The combined efficiency of the selection requirements to inelastic WIMP signals is estimated at $(93.6\pm0.7)\%$, determined using either simulated events or calibration data.
The dominant efficiency loss is due to the requirement of only a single S1 in an event, which has an acceptance of $95\%$ as measured with calibration data.

Correction factors are applied to the measured event quantities to account for systematic variations throughout the detector volume~\cite{Aprile:2019bbb}.
The most significant are corrections for the spatially-varying light collection efficiencies for the ionisation and scintillation signals, and the attenuation of the ionisation signal from charge loss due to electron attachment onto electronegative impurities in the liquid xenon.
An additional correction accounts for inhomogeneity in the electric fields.
The corrections transform the S1 signal into cS1, and the S2 into cS2.
As the S2 signal is generated only a few cm below the top PMT array, the response of this set of photosensors is significantly non-uniform.
Thus, only the S2 light seen by the bottom PMT array (\cstwob{}) is used for energy reconstruction.
The analysis is performed with the corrected quantities cS1 and \cstwob{}.

\subsection{Expected signal}
\label{sec:signal}

The response of XENON1T to an inelastically-scattering WIMP is modelled using a combined data-driven and Monte-Carlo-driven procedure.
The two energy depositions -- the NR from the collision and ER from the subsequent gamma ray --  are considered independently.
The complete response of XENON1T to an inelastic scatter is the sum of these two models.

The response to the NR is modelled analogously to previous elastic scattering analyses of XENON1T data.
The mass-dependent WIMP spectra of~\cite{Baudis:2013bba} are used together with the XENON1T response model described in~\cite{Aprile:2019dme}, based on a fit to calibration events induced by neutron sources.
Since the inelastic scatter is detected as a single event, we do not apply detector efficiencies to the separate signal components, but to the final combined signal model.

\begin{figure}
    \centering
    \includegraphics[width=246.0pt]{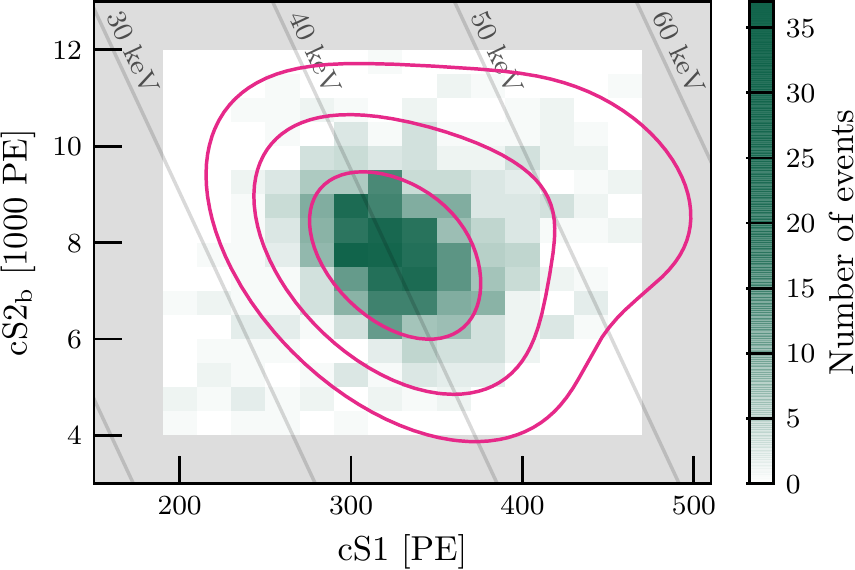}
    \caption{Model for inelastic scattering of neutrons (pink $1\sigma$, $2\sigma$ and $3\sigma$ contours) compared to calibration data with the neutron generator deployed. The density of calibration data is represented by the shading inside the $2\sigma$ contour, while outside individual events in the calibration data are shown as green dots. The white unshaded region indicates the region in which the fit is performed. Energy contours are shown as grey lines.}
    \label{fig:signalng}
\end{figure}

The ER signal is assumed to be well described by a two-dimensional (2D) Gaussian function in the plane of cS1 and \cstwob~\cite{Aprile:2020yad}.
The total (ER plus NR) response model for inelastic neutron scattering is fit to calibration data from a D-D neutron generator~\cite{Lang:2017ymt}, using a binned likelihood maximisation procedure, to extract the five parameters needed to describe this Gaussian.
The NR response is fixed for the fit, and determined using the method described above for WIMPs.
A background of pure ER events is present in this calibration data, predominantly due to decays of \lead.
We therefore fit an additional component, whose shape is assumed to be the same as the ER band in \rnzero{} calibration data (see section~\ref{sec:backgrounds}).
The best fit model for inelastic neutron scattering is shown in figure~\ref{fig:signalng} along with the neutron generator calibration data used for the fit.
The goodness-of-fit p-value is 0.22, computed from the likelihood compared to its distribution for toy data.

The ER model produced with these best-fit parameters is then combined with the NR model for a particular WIMP mass.
After considering the selection efficiencies, this gives the total expected response to such a WIMP inelastically scattering in XENON1T.

\subsection{Backgrounds}
\label{sec:backgrounds}

Four primary backgrounds contribute to the signal region of this analysis. These are decays of $^{214}$Pb (a daughter of \rntwo), residual contamination of the calibration isotope \krthree, and peaks from \xeonetwofour{} and \iodine~\cite{XENON:2019dti}.
Additional contributions from elastic scattering of solar neutrinos off electrons, decays of \krfive, $2\nu 2\beta$ decays of $^{136}$Xe, and Compton scatters of gammas emitted by detector components have rates which are smaller by at least one order of magnitude and are not considered~\cite{Aprile:2015uzo}.

\begin{figure}
    \centering
    \includegraphics[width=\linewidth]{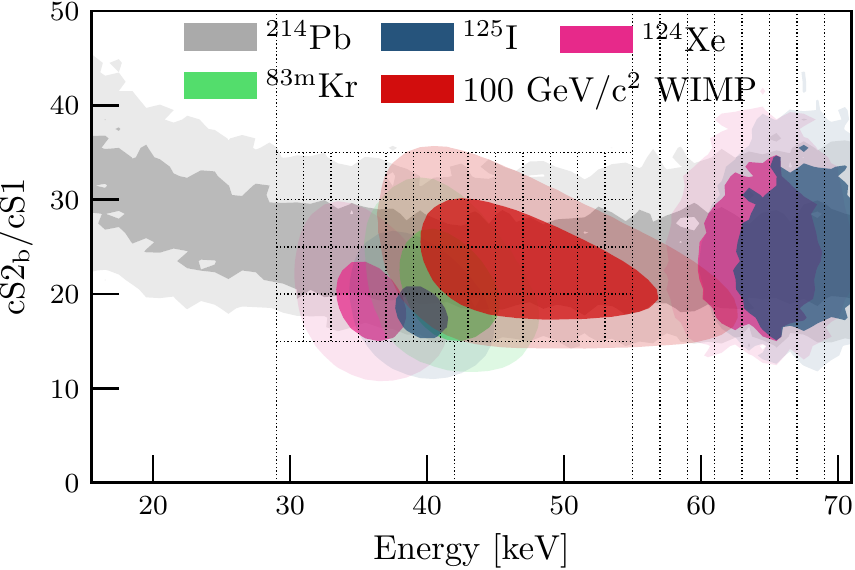}
    \caption{Region of interest showing the four nominal background models: \lead{} in grey, \krthree{} in green, \xeonetwofour{} in pink and \iodine{} in blue; and the expected signal for a $100\1{\gev}$ WIMP in red. In each case the darker and lighter shading shows the $1\sigma$ and $2\sigma$ region, respectively. The \cstwob{}/cS1 distribution for \lead{} is used to visualise the \dec-KK and EC-K peaks, which are only modelled in energy. The dotted lines indicate the boundaries of the bins used to perform the likelihood fit.}
    \label{fig:binning_and_models}
\end{figure}

The backgrounds presented here are modelled in the parameter space described by the ratio \cstwob{}/cS1 and energy.
Although ER and NR events have different energy scales, we use the ER-equivalent energy.
The $1\sigma$ and $2\sigma$ contours of each model are shown along with the expected signal for a $100\1{\gev}$ WIMP in figure~\ref{fig:binning_and_models}.

\subsubsection{\texorpdfstring{\lead}{Pb-214}}

The isotope \rntwo{} continuously emanates from all detector materials due to trace contamination of $^{238}$U~\cite{Aprile:2020vmn}.
Itself a noble element, radon is unaffected by purification techniques which target electronegative impurities such as oxygen and water.
Its half-life of $3.8\1{days}$ is sufficiently long for it to mix uniformly throughout the active volume~\cite{Aprile:2017fhu}, where it decays first to $^{218}$Po (half-life $3.1\1{min}$) then \lead{} (half-life $27\1{min}$).
Of the decays of \lead, $11\%$ are directly to the ground state of $^{214}$Bi and have no accompanying gamma emission~\cite{Wu:2009lpp}.
These ``naked'' beta decays can therefore deposit any energy up to the Q-value of $1.02\1{MeV}$, and some fraction of these fall within the energy region of interest in this analysis.

The spectral shape of this background is modelled using Monte Carlo simulations.
The \cstwob{}/cS1 distribution is based on \rnzero{} calibration data with decays of $^{212}$Pb.

\subsubsection{\texorpdfstring{\krthree}{Kr-83m}}

An isotope commonly used for calibration is \krthree{}, which is regularly injected into the TPC, where it spreads throughout the active region~\cite{Manalaysay:2009yq,Kastens:2009pa,Akerib:2017eql}.
A low rate of \krthree{} -- on average $10^{-4}$ times the rate during calibrations -- persisted in the detector even considerably after being injected, possibly due to trace contamination from its parent isotope $^{83}$Rb in the xenon purification loop.

The nature of this background is readily verifiable -- \krthree{} decays in a two-step process with an intermediate half-life of $157\1{ns}$~\cite{McCutchan:2015vcl}.
We can therefore select some events where the two S1 signals are separately identified.
Using events where the time between the two S1s is between 600 and $800\1{ns}$ (closer S1s are not always separately identified), we calculate the total rate of \krthree{} in the detector.
We compare the fraction of events which are tagged in the same way in \krthree{} calibration data, to the fraction which are accepted by the data quality criteria used for the inelastic scattering search.
Scaling by this fraction, we predict an average rate of $(18.2\pm1.0)\1{day^{-1}}$ over the full science run, decreasing from an initial rate around $60\1{day^{-1}}$ with the expected $86\1{day}$ half-life of the $^{83}$Rb parent.

Events where the two constituent decays are unresolvable result in a single signal at $41\1{keV}$, which lies in the energy region of interest.
Since most of the energy released is in the form of IC and Auger electrons, the ratio \cstwob/cS1 is similar to that of \lead.
We use \krthree{} calibration data, after applying the analysis selection criteria, to model the distribution of this background.

\subsubsection{\texorpdfstring{\xeonetwofour{}}{Xe-124} \& \texorpdfstring{\iodine}{I-125}}

The ER background stemming from \xeonetwofour{} and \iodine{} is modelled in two ways.
Each of these two isotopes provides two peaks in the region of interest, depending on which electron shells are involved in the decay.
\xeonetwofour, decaying via two-neutrino double electron capture (\dec), produces peaks at $64.3\1{keV}$ ($36.7\1{keV}$) with branching ratios $\sim76\%$ ($\sim23\%$) via capture from the KK (KL) electron shells (here denoted \dec-KK and \dec-KL, respectively)~\cite{Doi:1991xf}.
\iodine{} is produced by neutron capture on \xeonetwofour{} primarily during neutron calibrations, and decays via electron capture (EC). This produces peaks at $67.3\1{keV}$ ($40.4\1{keV}$) with branching ratios $80.1\%$ ($15.6\%$) via capture from the K (L) shell (here denoted EC-K and EC-L, respectively)~\cite{Be:2011aa}.

The models for these backgrounds are created in a different manner than for the previous two, since no directly comparable calibration data is available.
The \dec-KK and EC-K events lie relatively far from the signal region, so a model consisting of one-dimensional Gaussian peaks in energy is sufficient.

For the \dec-KL and EC-L decays, a one-dimensional model would have a significant adverse impact on the sensitivity.
In the following we describe the production of the two-dimensional model for \dec-KL; the model for EC-L is produced analogously.
The two-dimensional model is formed by scaling the light and charge yields of the \krthree{} calibration peak.
The contributions from X-rays and Auger electrons to the total energy deposition of \dec-KL decays are determined using RELAX~\cite{epics2017}.
The NEST package~\cite{Szydagis_2011} is then used to simulate the light and charge yields of liquid xenon to pure betas and pure gammas at the total energy deposition.
These are linearly combined based on their fractional contributions to estimate the total light and charge yield for \dec-KL.
The predicted light and charge yields for \krthree{} decays are directly obtained from NEST.
To estimate the cS1 (\cstwob{}) distribution for \dec{}-KL, the distribution observed in \krthree{} calibration data is scaled by the ratio between the \dec{}-KL and \krthree{} light (charge) yields.
The average charge yields of \dec{}-KL and EC-L decays are lower than those of \krthree{} and \lead{} due to the significant fraction of the energy released as gamma rays.

\subsection{Statistical interpretation}
\label{sec:stats}

For each WIMP mass considered between $20\1{\gev{}}$ and $10\1{TeV\!/c^2}$, a binned profiled likelihood is used to constrain the cross-section of inelastic WIMP interactions.
The binning structure used to evaluate the likelihood, shown in figure \ref{fig:binning_and_models}, is chosen in order to optimise sensitivity while preserving asymptotic properties of the likelihood \cite{Wilks:1938dza}.
This explains the larger bins used in the tail regions of the background, at high and low values of \cstwob{}/cS1.
Specifically, we ensure that a minimum of five background events are expected in every bin.
At energies above $55\1{keV}$, where \dec{}-KK and EC-K backgrounds are present, bins are only a function of energy, since these peaks are not modelled in two dimensions.
Events between 15 and $29\1{keV}$ are taken together to constrain the \rntwo{} rate.
This is taken into account using as a single ancillary likelihood term that is combined with the binned likelihood.

The data-derived backgrounds from \rntwo{} and \krthree{} have an uncertainty on the fraction of events expected in each bin due to the finite statistics of the calibration data used to create the model.
These statistical uncertainties can be treated with a simultaneous fit to the calibration and the science data.
We follow the method described in~\cite{Barlow:1993dm} which makes it possible to incorporate this into the fitting procedure in a computationally efficient way.

During the operation of XENON1T, a very high number of events from the decay of \krthree{} were recorded during the regular calibrations.
Therefore, in every bin, the statistical uncertainty arising from \krthree{} is never more than $0.4\%$ of that from the \rntwo{} model and can thus be neglected.
The effect of statistical fluctuations in the Monte-Carlo data used to produce the \rntwo{} model is also ignored -- at any given energy the statistics is between 700 and 800 times greater than the \rnzero{} calibration data.
This means that the simultaneous fit is only to one calibration source (\rnzero{}) in addition to the science data.
In this case the method of~\cite{Barlow:1993dm} becomes analytic, and no additional degrees of freedom must be introduced into the minimisation routine compared to fitting the science data alone.

Uncertainties on the mean energy and resolution of the \dec{}-KK and EC-K peaks are incorporated as independent nuisance parameters in the fit, each with a Gaussian constraint term.
The uncertainty on the charge yield for the \dec{}-KL and EC-L peaks is also included with a Gaussian constraint.
To constrain these parameters, the charge yield in the ER band at the energy of the \krthree{} peak is compared to the yield predicted by scaling the \krthree{} peak itself using the same procedure as for producing the models, described above.
The difference between predicted and measured charge yields in this case is taken as an uncertainty on the predicted yields for the \dec{}-KL and EC-L peaks.
Because the physical processes behind each peak are very similar, we assume that the uncertainties are correlated and a single nuisance parameter is used to vary the charge and light yields for both peaks simultaneously.

\begin{figure*}[p]
    \centering
    \includegraphics[width=510pt]{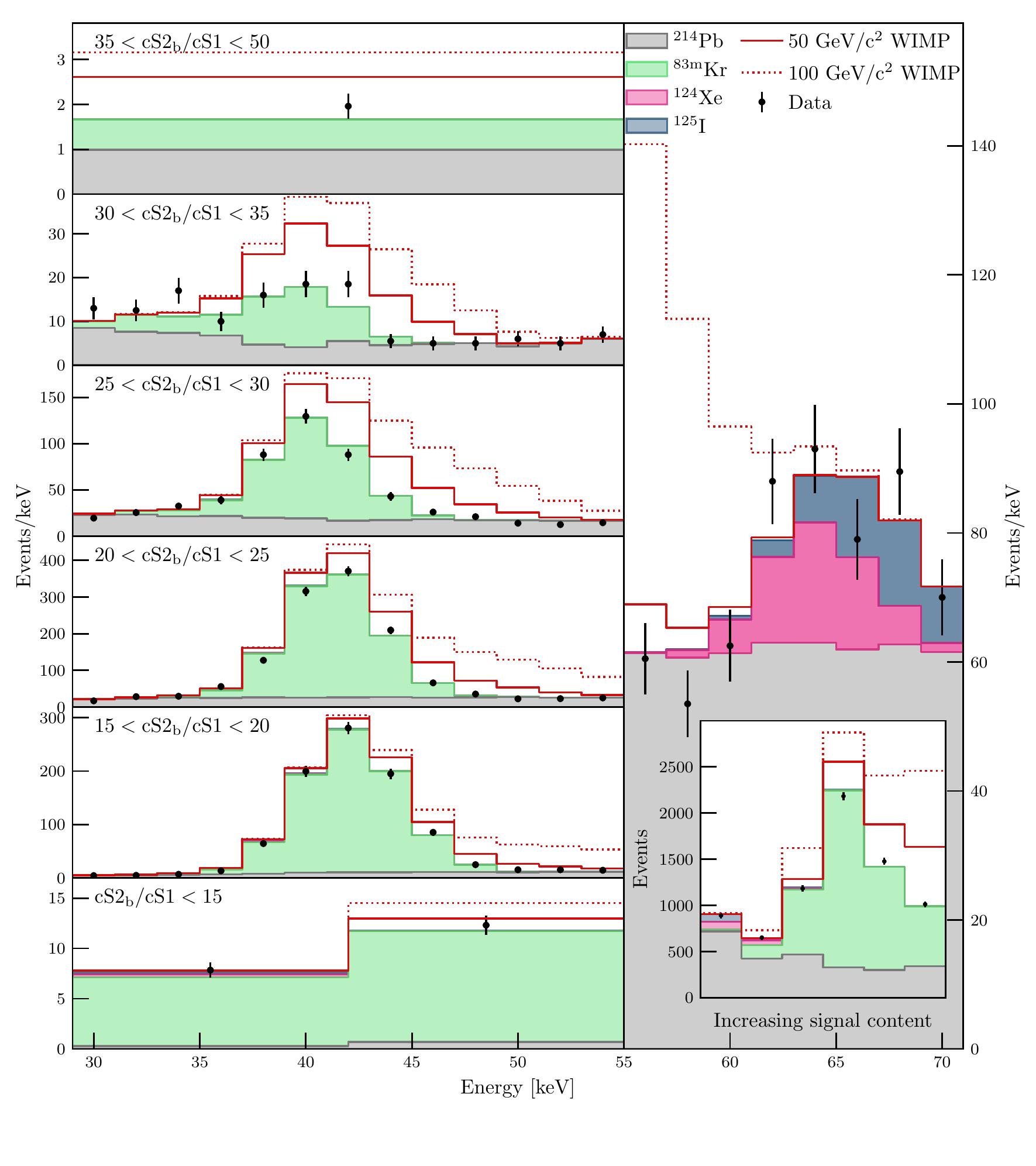}
    \caption{Observed events in the $0.83\1{t}$ exposure, shown as error bars. Between 29 and $55\1{keV}$, the six histograms correspond to the six \cstwob{}/cS1 bins as shown in figure \ref{fig:binning_and_models}. Above $55\1{keV}$ only a single \cstwob{}/cS1 bin is shown, as used for the analysis. The single bin from 15 to $29\1{keV}$ is not shown, since it is instead used to constrain the background rate in the rest of the analysis region. The shaded areas show the estimated background in each bin from \lead{} (grey), \krthree{} (green), \xeonetwofour{} (pink) and \iodine{} (blue), based on a background-only best fit to the science data. The solid (dotted) red line shows the signal expected from a $50\1{\gev{}}$ ($100\1{\gev}$) WIMP scattering inelastically, with a cross-section of $10^{-37}\1{cm^2}$, on top of the background estimation. The inset shows a more abstract representation of the same data (from the full region of interest). Analysis bins are ordered according to the ratio between the expected number of signal events from a $50\1{\gev}$ WIMP and the expected number of background events and then grouped into the six shown here. An upwards fluctuation is present in the right-hand bins, explaining the origin of the upwards fluctuation of the limit within the expected band.}
    \label{fig:signalarranged}
\end{figure*}

\begin{table}
    \caption{Overview of parameters with a Gaussian constraint in the likelihood. The effective efficiency has a mass-dependent uncertainty, varying between $0.936\pm0.033$ for $30\1{\gev}$ WIMPs and $0.936\pm0.014$ for $10\1{TeV\!/c^2}$ WIMPs.}
    \label{tab:likelihoodparams}
    \begin{ruledtabular}

    \begin{tabular}{l@{}D{+}{\,\pm\,}{5,10}@{}l}
        Parameter & \multicolumn{1}{l}{\;\;Constraint} \\ \colrule
        Relative \rntwo{} rate & 1.00 + 0.03 \\
        Mean energy of \dec{}-KK peak & (64.3 + 0.6)\1{keV} \\
        Resolution of \dec{}-KK peak & (2.6 + 0.3)\1{keV} \\
        Mean energy of EC-K peak & (67.3 + 0.5)\1{keV} \\
        Resolution of EC-K peak & (2.8 + 0.5)\1{keV} \\
        Variation in KL/L charge yield & (0.0 + 0.7)\1{e^-/keV} \\ 
        Effective efficiency & \multicolumn{1}{l}{Mass dependent} \\
        Mean \cstwob{} of inelastic ER signal & (7320 + 40)\1{PE} \\
    \end{tabular}
    \end{ruledtabular}
\end{table}

The effect on the signal model of eighteen parameters is considered.
These are the five parameters needed to describe the 2D Gaussian part of the model and the thirteen parameters from~\cite{Aprile:2019dme} which are relevant to the NR model.
The importance of each of these is assessed by determining how much the sensitivity changes when that parameter is varied by $\pm 1\sigma$ with the remaining parameters fixed to their nominal values.
The mean \cstwob{} of the ER part of the signal is found to be the most important, especially at low WIMP masses ($3.4\%$ effect at $20\1{\gev{}}$), and this parameter is included in the likelihood fit with a Gaussian constraint.
The others are combined into a single uncertainty on the signal rate, calculated as the sum in quadrature of their effect on the sensitivity.
This is also combined with the uncertainty on the efficiency of the selection criteria to form a single nuisance parameter which we term the ``effective efficiency''.
A summary of the constrained parameters is shown in Table~\ref{tab:likelihoodparams}.

\section{Results and discussion}
\label{sec:results}

The distribution of events observed in the 0.83 tonne-year exposure after unblinding is shown in figure~\ref{fig:signalarranged}.
A total of 7392 events are observed.
The number of events expected from each source of background, for the best fit models, and the best fit number of signal events, is detailed in table~\ref{tab:evs_per_source}.
Based on the profiled likelihood analysis, no significant evidence is found for spin-dependent inelastic WIMP-nucleon scattering in the search data.
A pre-unblinding decision was made to report only an upper limit if the discovery significance p-value was less than $0.003$ ($3\sigma$).
The highest significance is for $50\1{\gev}$ WIMPs, where the p-value for the background-only hypothesis is 0.023, from the log likelihood ratio.
A $90\%$ confidence level upper limit is therefore placed on the cross section.
Under the background-only scenario, we obtain a goodness-of-fit p-value of 0.055, by comparing the $\chi^2$ goodness of fit to its distribution for simulated data.

\begin{figure}[t]
    \centering
    \includegraphics[width=246.0pt]{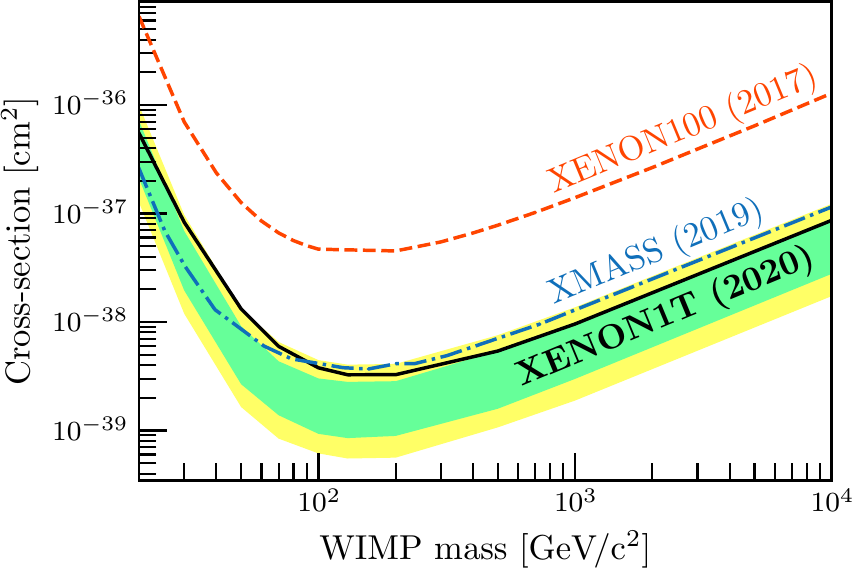}
    \caption{Upper limit, at $90\%$ CL, on the cross-section for inelastic scattering of WIMPs off xenon nuclei. The expected range at $1\sigma$ and $2\sigma$ is shown in green and yellow, respectively. Our result is compared to upper limits obtained by the XMASS collaboration~\cite{Suzuki:2018xek} (blue dot-dashed line) and XENON100~\cite{Aprile:2017ngb} (orange dashed line).}
    \label{fig:limit}
\end{figure}

The result is shown in figure~\ref{fig:limit}, along with the expected $1\sigma$ and $2\sigma$ range of upper limits.
We set the strongest upper limit for WIMP masses greater than $100\1{\gev}$, reaching $3.3 \times 10^{-39}\1{cm^2}$ at a WIMP mass of $130\1{\gev}$.
The sensitivity to lower-mass WIMPs (up to about $30\1{\gev}$) is severely affected due to the similarity between their signal shape and the \krthree{} background.
For the lowest masses considered, this effect weakens the sensitivity by approximately a factor of three; above $130\1{\gev}$ the effect is approximately $10\%$.
Our limit is weaker than the upper $1\sigma$ quantile of the expected range of limits for all WIMP masses.
This is due to a slight over-fluctuation in the bins with the most sensitivity to WIMPS, as seen in the inset of figure \ref{fig:signalarranged}.
We also compare the limit to that reported by the XMASS collaboration in~\cite{Suzuki:2018xek}, obtained from a 0.72 tonne-year exposure of XMASS-I.

\begin{table}
    \caption{Best fit expectation values for the number of events from $50\1{\gev{}}$ WIMPs and each background, under the likelihood's best fit scenario. ``All bins'' refers to the entire region of interest (15--71 keV, $0\leq \cstwob{}/\mathrm{cS1}\leq 50$), and ``signal region'' refers to the bins containing 95\% of the signal model.}
    \label{tab:evs_per_source}
    \begin{ruledtabular}
    \begin{tabular}{crr}
        Source & \multicolumn{1}{r}{All bins} & \multicolumn{1}{r}{Signal region}\\ \colrule
        \lead{} & 2546 & 1084\\
        \krthree{} & 4529 & 4164\\
        \xeonetwofour{} & 159 & 6 \\
        \iodine{} & 114 & 12 \\
        WIMP & 28 & 26 \\ \colrule
        Total & 7376 & 5292 \\
        Data & 7392 & 5320\\
    \end{tabular}
    \end{ruledtabular}
\end{table}

These results complement previously reported searches for elastic WIMP scattering~\cite{Aprile:2018dbl,Aprile:2019dbj}.
The sensitivity is lower for most WIMP-scattering scenarios.
However, a positive detection of inelastic scattering would place stronger constraints on the properties of the dark matter than a detection of elastic scattering alone.

Next-generation experiments including XENONnT, which is currently being commissioned, are expected to have an ER background rate around five times lower~\cite{Aprile:2020vtw}.
In addition, a much larger sensitive region will mean that larger exposures can quickly be obtained.
This will allow even smaller cross-sections of inelastic scattering to be probed.

\section{Acknowledgements}
We gratefully acknowledge support from the National Science Foundation, Swiss National Science Foundation, German Ministry for Education and Research, Max Planck Gesellschaft, Deutsche Forschungsgemeinschaft, Helmholtz Association, Netherlands Organisation for Scientific Research (NWO), Weizmann Institute of Science, ISF, Fundacao para a Ciencia e a Tecnologia, Région des Pays de la Loire, Knut and Alice Wallenberg Foundation, Kavli Foundation, JSPS Kakenhi in Japan and Istituto Nazionale di Fisica Nucleare. This project has received funding or support from the European Union’s Horizon 2020 research and innovation programme under the Marie Sklodowska-Curie Grant Agreements No. 690575 and No. 674896, respectively. Data processing is performed using infrastructures from the Open Science Grid, the European Grid Initiative and the Dutch national e-infrastructure with the support of SURF Cooperative. We are grateful to Laboratori Nazionali del Gran Sasso for hosting and supporting the XENON project.

\newpage

\end{document}